\documentclass[12pt,notitlepage,aps,prc,nofootinbib,superscriptaddress,showpacs,showkeys,preprintnumbers,floatfix,byrevtex]{revtex4-1}

\usepackage{soul}										
\setstcolor{red}	                                  

\usepackage{epsfig}
\usepackage{amsmath,amsthm,amssymb}
\usepackage{multirow}
\usepackage{graphicx}
\usepackage[usenames,dvipsnames,svgnames,table]{xcolor}
\DeclareMathAlphabet{\mathpzc}{OT1}{pzc}{m}{it}

\newcommand{\Ex}[2]{\ifmmode{#1\times10^{#2}}\else{$#1\times10^{#2}$}\fi}

\newcommand{\ud}{\mathrm{d}}

\newcommand{\nn}{\nonumber}

\newcommand{\fig}[1]{Fig.\,\ref{#1}}

\newcommand{\eqn}[1]{Eq.\,\eqref{#1}}

\newcommand{\unit}[1]{\,\mathrm{#1}}
\newcommand{\fm}{\,\mathrm{fm}}

\newcommand{\MeV}{\,\mathrm{MeV}}

\newcommand{\atom}[2]{\ifmmode{{}^{#2}\text{#1}}\else{${}^{#2}\text{#1}$}\fi}

\begin{document}

\title{Effects of pairing correlations on the neutron skin thickness and the symmetry energy}

\author{Soonchul Choi}
\affiliation{Department of Physics and OMEG Institute, Soongsil University, Seoul 06978, Korea}

\author{Ying Zhang}
\affiliation{Department of Physics, School of Science, Tianjin University, Tianjin 300072, China}

\author{Myung-Ki Cheoun}
\email{cheoun@ssu.ac.kr}
\affiliation{Department of Physics and OMEG Institute, Soongsil University, Seoul 06978, Korea}

\author{Youngshin Kwon}
\affiliation{Research Institute of Basic Science, Korea Aerospace University, Goyang 412-791, Korea}

\author{Kyungsik Kim}
\affiliation{School of Liberal Arts and Science, Korea Aerospace University, Goyang 412-791, Korea}

\author{Hungchong Kim}
\affiliation{Research Institute of Basic Science, Korea Aerospace University, Goyang 412-791, Korea}

\date{\today}

\begin{abstract}
We investigated effects of pairing correlations on the neutron skin thickness and the symmetry energy of finite nuclei. In this calculation we used Hartree-Fock-Bogoliubov (HFB) method with Skyrme forces and effective pairing interactions. The results have been compared with available experimental data, Hartree-Fock (HF) results as well as the predictions by droplet model (DM). Finally, our discussion was extended to study of the pairing interaction in nuclear matter. Roles of isospin $T=0$ pairing in the nuclear matter were also discussed.
\end{abstract}


\maketitle

\section{Introduction}
The study of symmetry energy and its density-dependence is a topic of great interest in nuclear physics and astrophysics. Its importance emerges out of understanding the structure of neutron-rich nuclei, various observational data of neutron stars and recent heavy ion collision data because the isospin- and density-dependence of the equation of state (EoS) in nuclear matter can be characterized by the symmetry energy~\cite{Li:2008gp}. The nuclear EoS, which explicates the energy per nucleon of asymmetric nuclear matter, can be expanded as a function of neutron-proton asymmetry, $\delta = (N-Z)/A$ (or equivalently $\delta = (\rho^{}_n - \rho^{}_p)/\rho$), at a given density $\rho$
\begin{equation}
	\mathcal{E}(\rho, \delta) =\mathcal{E}(\rho, 0) + \mathcal{E}^{}_{sym}(\rho)\delta^2 + \cdots~.
 \label{eq1}
\end{equation}
The coefficient of the quadratic in \eqn{eq1} is the nuclear symmetry energy that can be approximated to the energy difference between the pure neutron matter (PNM) and symmetric nuclear matter (SNM).  Since most of the nuclear models available these days are adjusted to the data of the binding energy of finite nuclei, they agree on the value of the symmetry energy around the saturation density $\rho^{}_0$.  However, when it comes to its density-dependence, there is a strong model dependence. In order to characterize the density dependence, one explores the symmetry energy around saturation density, expanding it with respect to the density $x = (\rho - \rho^{}_0)/(3\rho^{}_0)$
\begin{equation}
	\mathcal{E}^{}_{sym} = J + Lx +\frac{1}{2} K^{}_{sym} x^2 +  \cdots~,
 \label{eq2}
\end{equation}
where  $L$ and $K^{}_{sym}$ are the slope and the curvature parameters of nuclear symmetry energy at $\rho^{}_0$. The leading term, $J$, indicates the symmetry energy at $\rho^{}_0$ which values are predicted in the order of $30 \sim 35\MeV$. The 2nd coefficient, which is the slope parameter, $L (= 3 \rho_0 {  {\partial \mathcal{E}_{sym} ( \rho)    } \over {\partial \rho  }} |_{\rho = \rho_0}  )$, is related more or less to the pressure as follows
\begin{equation}
P (\rho, \delta) = \rho^2 ( {  { d \mathcal{E}  (\rho, 0) } \over {d \rho  }}   +  \delta^2 {  {d \mathcal{E}_{sym} ( \rho)    } \over {d \rho  }} )  ~,
\end{equation}
and still has a wide ambiguity ($20\sim100\MeV$)~\cite{Vinas:2013hua}.

Another simple parametrization of the symmetry energy is usually introduced in the interpretation of heavy ion collisions
\begin{equation}
	\mathcal{E}^{}_{sym} \simeq \mathcal{E}^{}_{sym}(\rho^{}_0)\left(\frac{\rho}{\rho^{}_0}\right)^\gamma = J(1+3x)^\gamma~,
 \label{eq3}
\end{equation}
from which the slope and the curvature can be represented as $L = 3J\gamma$ and $K^{}_{sym} = 9J\gamma(\gamma-1)$ \cite{Vinas:2013hua}. Many of nuclear experiments have been conducted to make a constraint on the density-dependence of the symmetry energy. The parameter $\gamma = 0.4 -1.05$ ($L = 88\pm 25\MeV$) is constrained by the isospin diffusion data in heavy ion collisions~\cite{Tsang:2008fd}, while $\gamma = 0.69$ ($L\sim 65\MeV$) is deduced from isotope ratios~\cite{Chen:2004si}. Discussion of nuclear collective motions can also give a unique chance to determine the incompressibility of the nuclear system, which is strongly related to the symmetry energy~\cite{Khan:2010mv,Piekarewicz:2006ip,Klimkiewicz:2007zz,Paar:2005fb,Sagawa:1999zz,Baran:2013gka}. For instance, the giant dipole resonance in heavy nuclei reflects the value of $\gamma = 0.5 - 0.65$~\cite{Trippa:2008gr}, and the Thomas-Fermi model results in $\gamma = 0.51$~\cite{Myers:1995wx}. Although the substantial progress has been made theoretically and experimentally, the density-dependence of the symmetry energy still remains uncertain and more accurate information is required to understand it.

Recently it has been regarded that the neutron skin thickness (NST), defined by the difference of the root-mean-square (rms) radii of protons and neutrons, is a conceivable clue for the symmetry energy~\cite{Typel:2001lcw,Centelles:2008vu,RocaMaza:2011pm,RocaMaza:2011wh}
\begin{equation}
	\Delta r^{}_{np} = \sqrt{\bigl<r^2_n\bigr>} - \sqrt{\bigl<r^2_p\bigr>}~.
 \label{eq4}
\end{equation}
The NST depends on the pressure of EoS in nuclear matter, and thus is related to the first derivative of the symmetry energy like the slop parameter, $L$~\cite{Typel:2001lcw,Alex00}. It can be extracted mostly through the antiprotonic measurements~\cite{Trzcinska:2001sy} and the parity-violating electron scattering~\cite{Horowitz:2012tj,Tsang:2012se,Kim:2013}. Neutrino-nucleus scattering in recent neutrino beam facilities could also be an alternative study for the NST~\cite{Patton:2012jr}.

It is noticeable that the formation of the neutron skin can be affected by the pairing correlation of nucleons, which is known to have a minor effect near the saturation density $\rho \approx \rho^{}_0$. However, in the surface of finite nuclei where the densities become lower than the saturation density, the contribution of the pairing correlations is no longer negligible~\cite{Khan:2010mv}. The pairing correlation should be also considered for the proper account of the symmetry energy because the effects of the isospin asymmetry are closely associated with the pairing correlations in the mean-field description of finite nuclei~\cite{Ring:1996qi,Neergard:2009rn}. For example, the mass-number dependence of the mass excess of $N=Z$ nuclei has a strong connection to the pairing correlations~\cite{Vogel:1998km}. It is thus worthwhile to discuss the effects of pairing correlation on the NST and the symmetry energy.

In the present work, we mainly employ the Hartree-Fock-Bogoliubov (HFB) method with Skyrme effective interactions to estimate the symmetry energy of various nuclei as a function of NST, concentrating on the influence of pairing correlations. In Section II, we first briefly review the definition of the symmetry energy and coefficient for finite nucleus, and its relation with the NST in the droplet model (DM).  Section III and IV  are dedicated to the main calculations and results given by the mean-field models with Skyrme functionals. In section V, we discuss the effect of the pairing correlations on infinite nuclear matter. Then this paper is wrapped up with a summary and conclusions in section VI.

\section{Neutron skin thickness and symmetry energy in droplet model}

Possible correlations between the NST ($\Delta r_{np}$) and the slop of symmetry energy, $L$, can be inferred from the DM~\cite{Centelles:2008vu,RocaMaza:2011pm,RocaMaza:2011wh}. In this model, the NST is given as by~\cite{Myers:1980NPA}
\begin{equation}
	\Delta r^{DM}_{np} = \sqrt{\frac{3}{5}} \,\left[ t - \frac{e^2 Z}{70J} \right]+\Delta r^{sw}_{np}~,
 \label{eq5}
\end{equation}
where the quantity $t$ is a distance between the neutron and proton mean surface locations,  the second term in the bracket is due to the Coulomb repulsion. The bulk contribution $t$ is given as
\begin{equation}
	t = \frac{3 }{2}r^{}_0 \frac{J}{Q}\frac{1}{1+x_A}(\delta-I_C),
	\quad\text{with}\quad x^{}_A = \frac{9J}{4Q}A^{-1/3}~,
\label{eq:t0}
\end{equation}
where $J$ is the leading term of the symmetry energy defined before, $Q$ is called the surface stiffness coefficient, which can be extracted from semi-infinite nuclear matter calculations~\cite{Warda:2009tc}, and $I^{}_C = e^2 Z/(20JR)$ with $R = r^{}_0 A^{1/3}$.
The term $\Delta r^{sw}_{np}$ is a correction caused by the difference in the surface widths $b_n$ and $b_p$ of the neutron and proton density profiles as
\begin{equation}
	\Delta r^{sw}_{np} = \sqrt{\frac{3}{5}}\,\frac{5}{2R} \left( b^2_n - b^2_p \right).
	\label{eq:skin-sw-b}
\end{equation}
It can be fitted as the following ansatz
\begin{equation}
	\Delta r^{sw}_{np} = \left( 0.3\, \frac{J}{Q} +c \right) \delta~,
	\label{eq:skin-sw-c}
\end{equation}
where $c = -0.05$ or $0.07\fm$ is parameterized to give a band region for different nuclear model calculations~\cite{Warda:2009tc}.
In the DM, the symmetry energy contribution to a finite nucleus with a mass $A$ is given by~\cite{Myers:1977book}
\begin{equation}
	E^{DM}_{sym} = a^{DM}_{sym}(A)\,\left( \delta+x^{}_A I^{}_C \right)^2 A
	\label{eq:EsymDM}
\end{equation}
where
\begin{equation}
	a^{DM}_{sym}(A) = \frac{J}{1+x^{}_A}
	\label{eq:asymDM}
\end{equation}
is the symmetry energy coefficient of the corresponding nucleus. Therefore, the bulk contribution $t$ to the neutron skin then can be written in terms of symmetric matter properties as follows
\begin{eqnarray}
 	t  &=&	 \frac{3 }{2}r^{}_0 \frac{a^{DM}_{sym}(A)}{Q}(\delta-I_C)  \label{eq:tasym1} \\
 	&=& \frac{2r_0}{3J}\left[ J-a^{DM}_{sym}(A)\right]A^{1/3}(\delta-I_C).
 	\label{eq:tasym2}
\end{eqnarray}

For a given mass number $A$, e.g., $^{208}$Pb, it is found universally in the mean-field calculation that the symmetry energy coefficient $a^{DM}_{sym}(A)$ equals the value of $\mathcal{E}^{}_{sym}(\rho)$ in Eq. \eqref{eq2} of asymmetry nuclear matter at the density $\rho=0.1$fm$^{-3}$.  With this relation, one can further reduce the bulk contribution $t$ in Eq.~\eqref{eq:tasym2} as
\begin{equation}
	t = \frac{2 r^{}_0}{3J} L \left[ 1+x\frac{K^{}_{sym}}{2L} \right] x A^{1/3} \left(\delta - I^{}_C\right)~,
	\label{eq6}
\end{equation}
which shows a linear correlation between the NST ($\Delta r^{DM}_{np} $) and the slope of the symmetry energy $L$~\cite{Centelles:2008vu}.
Equation ~\eqref{eq6} is different from the $t$ in Ref. \cite{Centelles:2008vu}, where $\epsilon = \frac{\rho_0-\rho}{3\rho_0}$ is used instead of $x=\frac{\rho-\rho_0}{3\rho_0}$.
On the other hand, for a given nuclear force, in which $Q$ can be determined from the corresponding semi-infinite nuclear matter calculations, one can also infer from Eq.~\eqref{eq:tasym1} a linear correlation between the NST ($\Delta r^{DM}_{np} $) and the symmetry energy coefficient $a^{DM}_{sym}(A)$ for an isotopic chain with different mass number $A$.  This could be seen in the figures shown in the Section IV.

Also for a given nuclear force with fixed $L$ and $K_{sym}$, one could infer from Eq.~\eqref{eq6} a linear relation between the NST ($\Delta r_{np}$) and the asymmetry parameter $\delta$, especially for heavier nuclei.  This relation is confirmed by the anti-protonic atom x-ray measurement~\cite{Trzcinska:2001sy}, and analyzed within the droplet model, Skyrme and relativistic mean-field models~\cite{Centelles:2008vu, Warda:2009tc}, which were also used to estimate the slope of the symmetry energy $L$.

\section{Isospin-dependence of neutron skin thickness in different Skyrme models}

In this section, we calculate the NST ($\Delta r_{np}$) in terms of the asymmetry parameter $\delta$, using HFB theory with 3 Skyrme functionals, SkI3, SLy4, and SVI, which have quite different density dependence (characterized by $L$ and $K_{sym}$) for the symmetry energy as shown in Ref.~\cite{Dutra:2012mb}.
The root mean square radius $\sqrt{\bigl<r^2_{n(p)}\bigr>}$ in Eq.~\eqref{eq4} is calculated by the self-consistent neutron (proton) density.
In the HFB calculations, we use the box-discretized method with a box size, $R^{}_{box} = 20\fm$, and the mesh size, $\ud r = 0.1\fm$. The angular momentum cut-off is taken as $l^{}_{max} = 12\hbar$. For the pairing interaction, we use the density-dependent delta interaction (DDDI) for the like-pairing field~\cite{Khan:2010mv,Ying17}
\begin{equation}
	G{(\mbox{\boldmath$r$})}_q = \frac{v_0}{2}\left[1-\eta\left(\frac{\rho^{}_q(\mbox{\boldmath$r$})}{\rho^{}_0}\right)^\alpha\right]\tilde{\rho}(\mbox{\boldmath$r$})~,\quad q=n~\text{or}~p~,
	\label{eq15}
\end{equation}
where $\tilde{\rho}$ is the pair density. The parameters in the DDDI are taken as $v^{}_0 = -458\unit{MeV/fm^{3}}$, $\eta=0.71$, $\alpha=0.59$, $\rho^{}_0=0.08$~fm$^{-3}$, which are determined for the Sn isotopes~\cite{Matsuo:2005vf}. The quasiparticle energy cut-off is $E^{}_{cut} = 60\MeV$.

\begin{figure}[t]
	\includegraphics[width=0.85\linewidth]{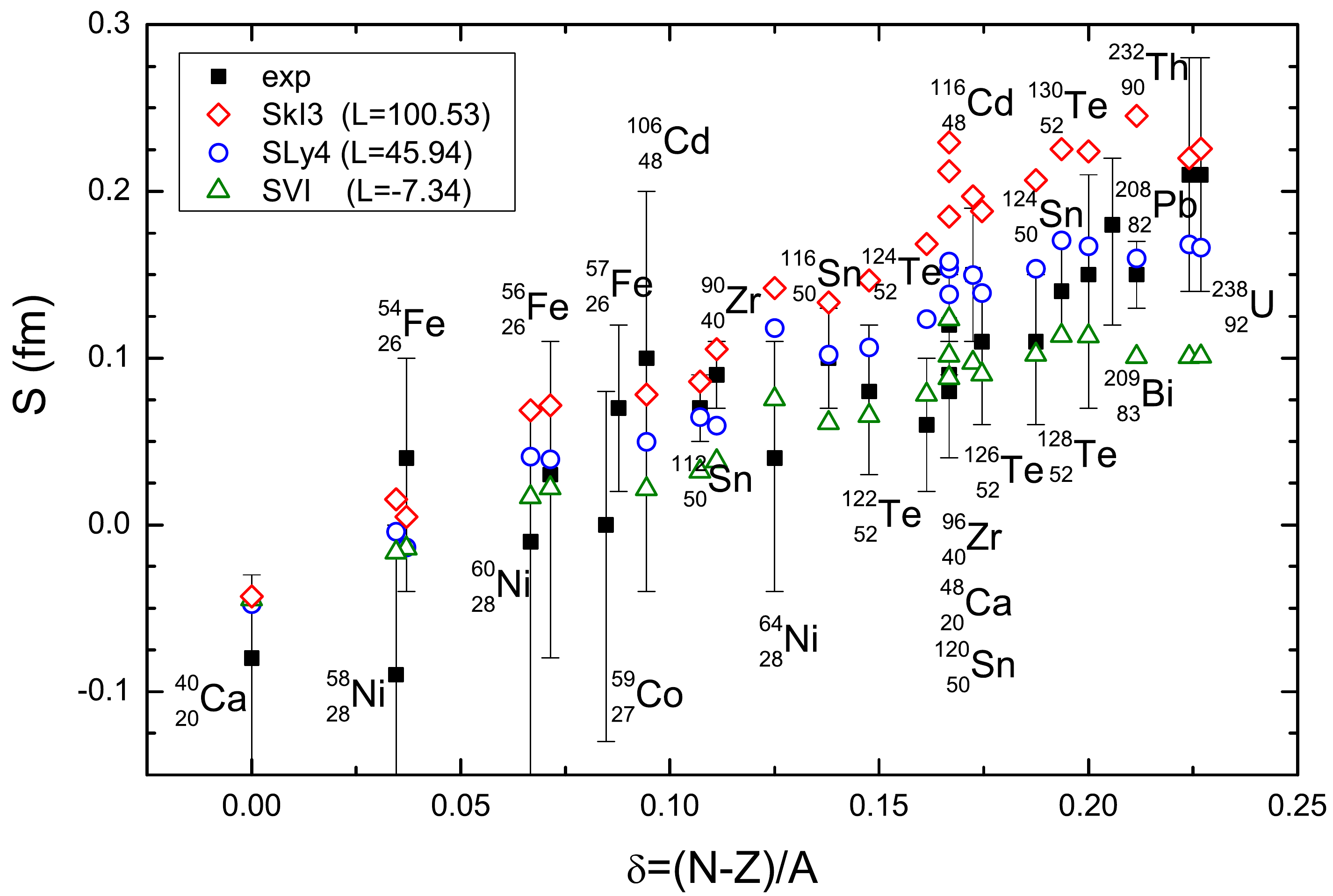}
	\caption{(Color online) Neutron skin thickness (S) calculated by the Skyrme Hartree-Fock-Bogoliubov (HFB) model with SkI3, SLy4, SVI interactions and density-dependent delta interaction (DDDI) for the pairing force, comparing with the anti-protonic measurement data~\cite{Trzcinska:2001sy}.}
	\label{fig1}
\end{figure}

Figure~\ref{fig1} shows our results of the NST ($S=\Delta r_{np}$), as a function of $\delta=(N-Z)/A$ compared with the experimental results deduced from the antiprotonic atom x-ray data~\cite{Trzcinska:2001sy}.
Our results are consistent with the data within uncertainties and comparable with the results in Refs.~\cite{Centelles:2008vu, Warda:2009tc}, although we include the shell and pairing effects in the full self-consistent calculations.  One can see that, even if the three Skyrme functionals have quite different $L$ and $K_{sum}$, a linear relation between the NST and the asymmetry parameter $\delta$ still holds to some extent.  However, the uncertainties of experimental data are still too large to deduce the $L$ value in the present model.

\section{Neutron skin thickness and symmetry energy with pairing effects}
In this section, we investigate the relation between the NST $\Delta r_{np}$ in Eq.~\eqref{eq4} and the symmetry energy $E_{sym}(A)$ or the symmetry energy coefficient $a_{sym}(A)$ for finite nuclei with different mass numbers by including the effects of pairing correlations. In order to calculate the symmetry coefficient $a_{sym}(A)$ for finite
nuclei microscopically from Skyrme functionals, we use the following recipe to rewrite the Skyrme functional as~\cite{Dong:2012ah}
\begin{equation}
 \begin{split}
 	\mathcal{H} &= \frac{1}{2}\hbar^2 \left( f^{}_n \tau^{}_n + f^{}_p \tau^{}_p \right) \\
 	&\quad + \left[ \frac{t^{}_0}{2}\left( 1+\frac{x^{}_0}{2} \right) +\frac{t^{}_3}{12}\left( 1+\frac{x^{}_3}{2} \right)\rho^\alpha \right]\rho^2 \\
 	&\quad + \left[ \frac{3\,t^{}_1}{16}\left( 1+\frac{x^{}_1}{2} \right) -\frac{t^{}_2}{16}\left( 1+\frac{x^{}_2}{2} \right) \right]\left(\nabla\rho\right)^2\\
 	&\quad - \left[ \frac{t^{}_0}{2}\left( x^{}_0+\frac{1}{2} \right) +\frac{t^{}_3}{12}\left( x^{}_3+\frac{1}{2} \right)\rho^\alpha \right] \\&\qquad\times\left( \rho^2_n+\rho^2_p \right) \\
 	&\quad - \left[ \frac{3\,t^{}_1}{16}\left( x^{}_1+\frac{1}{2} \right) +\frac{t^{}_2}{16}\left( x^{}_2+\frac{1}{2} \right) \right] \\&\qquad\times\left( \left(\nabla\rho^{}_n\right)^2+\left(\nabla\rho^{}_p\right)^2 \right) \\
 	&\quad + \frac{1}{16}\left[\left(t^{}_1-t^{}_2\right)\left(J^2_n+J^2_p\right) - \left(t^{}_1 x^{}_1+t^{}_2 x^{}_2\right)J^2\right] \\
 	&\quad + \frac{W^{}_0}{2}\left[ \mathbf{J}\cdot \nabla\rho+\mathbf{J}^{}_n\cdot \nabla\rho^{}_n + \mathbf{J}^{}_p\cdot\nabla\rho^{}_p \right] + \mathcal{H}^{}_c~,
 \end{split}
 \label{eq10}
\end{equation}
where
\begin{equation}
 \begin{split}
 	f^{}_{n, p} &= \frac{1}{m}+\frac{1}{\,\,2\hbar^2}\left[ t^{}_1\left(1+\frac{x^{}_1}{2}\right) +t^{}_2\left(1+\frac{x^{}_2}{2}\right) \right]\rho \\
 	&\quad - \frac{1}{\,\,2\hbar^2}\left[ t^{}_1\left(x^{}_1+\frac{1}{2}\right) - t^{}_2\left(x^{}_2+\frac{1}{2}\right) \right]\rho^{}_{n, p}~.
 \label{eq11}
 \end{split}
\end{equation}
If we assume $\rho^{}_n = \rho^{}_p = \frac{1}{2}\rho$ for symmetric nuclear matter, i.e.,
\begin{equation}
 \begin{split}
	f^{}_{n=p} &= \frac{1}{m}+\frac{1}{\,\,2\hbar^2}\left[ t^{}_1\left(1+\frac{x^{}_1}{2}\right) +t^{}_2\left(1+\frac{x^{}_2}{2}\right) \right]\rho \\
 	&\quad - \frac{1}{\,\,4\hbar^2}\left[ t^{}_1\left(x^{}_1+\frac{1}{2}\right) - t^{}_2\left(x^{}_2+\frac{1}{2}\right) \right]\rho~,
 \label{eq12}
 \end{split}
\end{equation}
the Hamiltonian is denoted as $\mathcal{H}^{}_0$. By excluding the Coulomb energy and the spin energy in $\mathcal{H} - \mathcal{H}^{}_0$, the density functional of the symmetry energy is given as
\begin{equation}
	\mathcal{H}^{}_{sym} = \mathcal{H}^{}_T + \mathcal{H}^{}_V+\mathcal{H}^{}_{grad}
 \label{eq13}
\end{equation}
with
\begin{equation}
 \begin{split}
	\mathcal{H}^{}_T &= \frac{\hbar^2}{2} \left( f^{}_n\tau^{}_n +  f^{}_p\tau^{}_p -2f^{}_{n=p}\tau^{}_{n=p} \right)~,\nn \\
	\mathcal{H}^{}_V &= -\left[ \frac{t^{}_0}{4}\left(x^{}_0+\frac{1}{2}\right) +\frac{t^{}_3}{24}\left(x^{}_3+\frac{1}{2}\right)\rho^\alpha\right]\rho^2\delta^2~,\\
	\mathcal{H}^{}_{grad} &= -\left[\frac{3\,t^{}_1}{32}\left(x^{}_1+\frac{1}{2}\right)+\frac{t^{}_2}{32}\left(x^{}_2+\frac{1}{2}\right)\right]\left[\nabla\left(\rho\delta\right)\right]^2~.
 \end{split}
\end{equation}

Then the total symmetry energy $E_{sym}(A)$ and the symmetry energy coefficient $a_{sym}(A)$ for a finite nuclei with the mass number $A$ are defined as
\begin{equation}
	E^{}_{sym}(A) = \int\mathcal{H}^{}_{sym}\ud V = a^{}_{sym}(A)\,\left( \delta+x^{}_A I^{}_C \right)^2A~.
 \label{eq14}
\end{equation}

\begin{figure}[h]
	\includegraphics[width=0.85\linewidth]{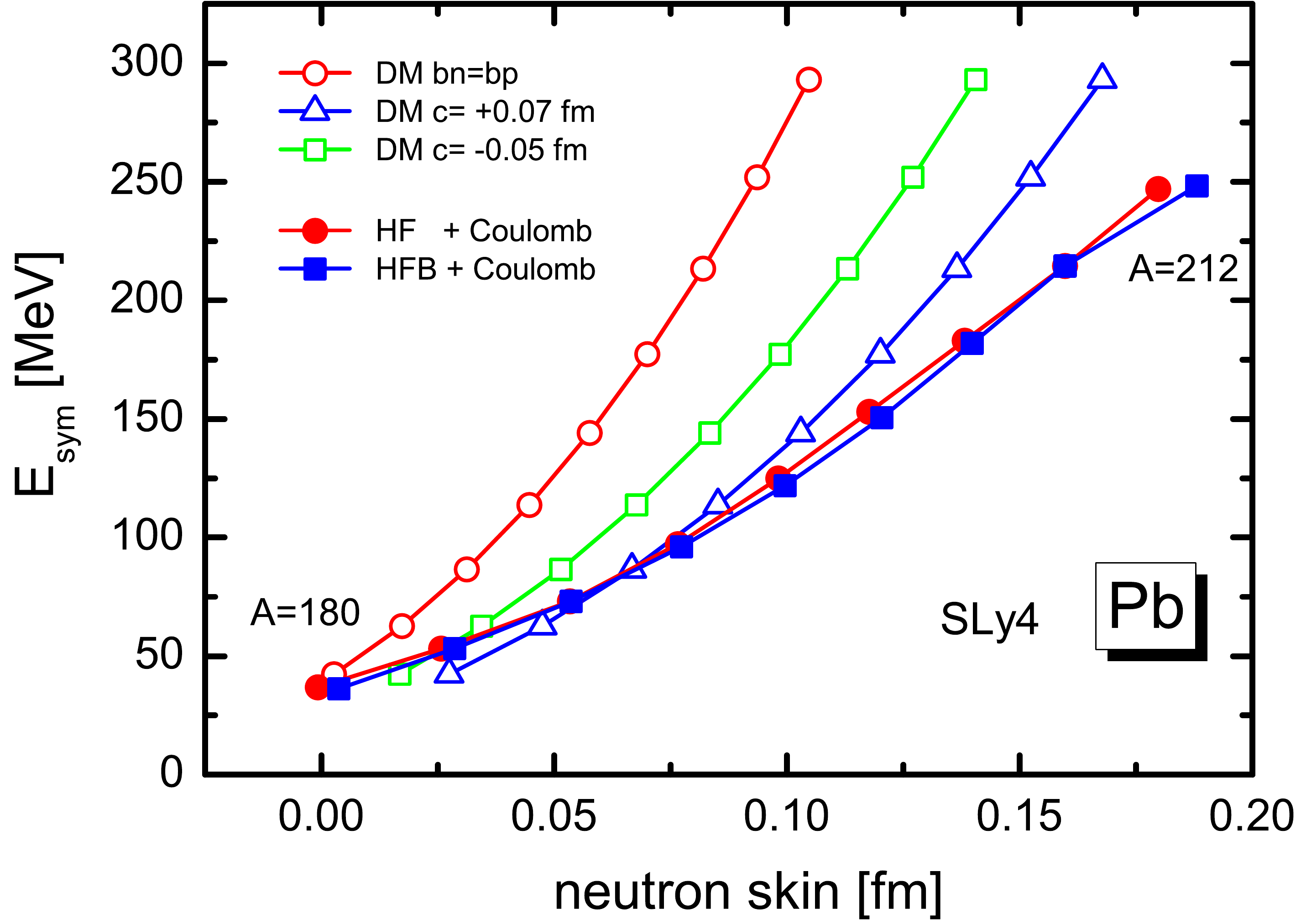}
	\includegraphics[width=0.85\linewidth]{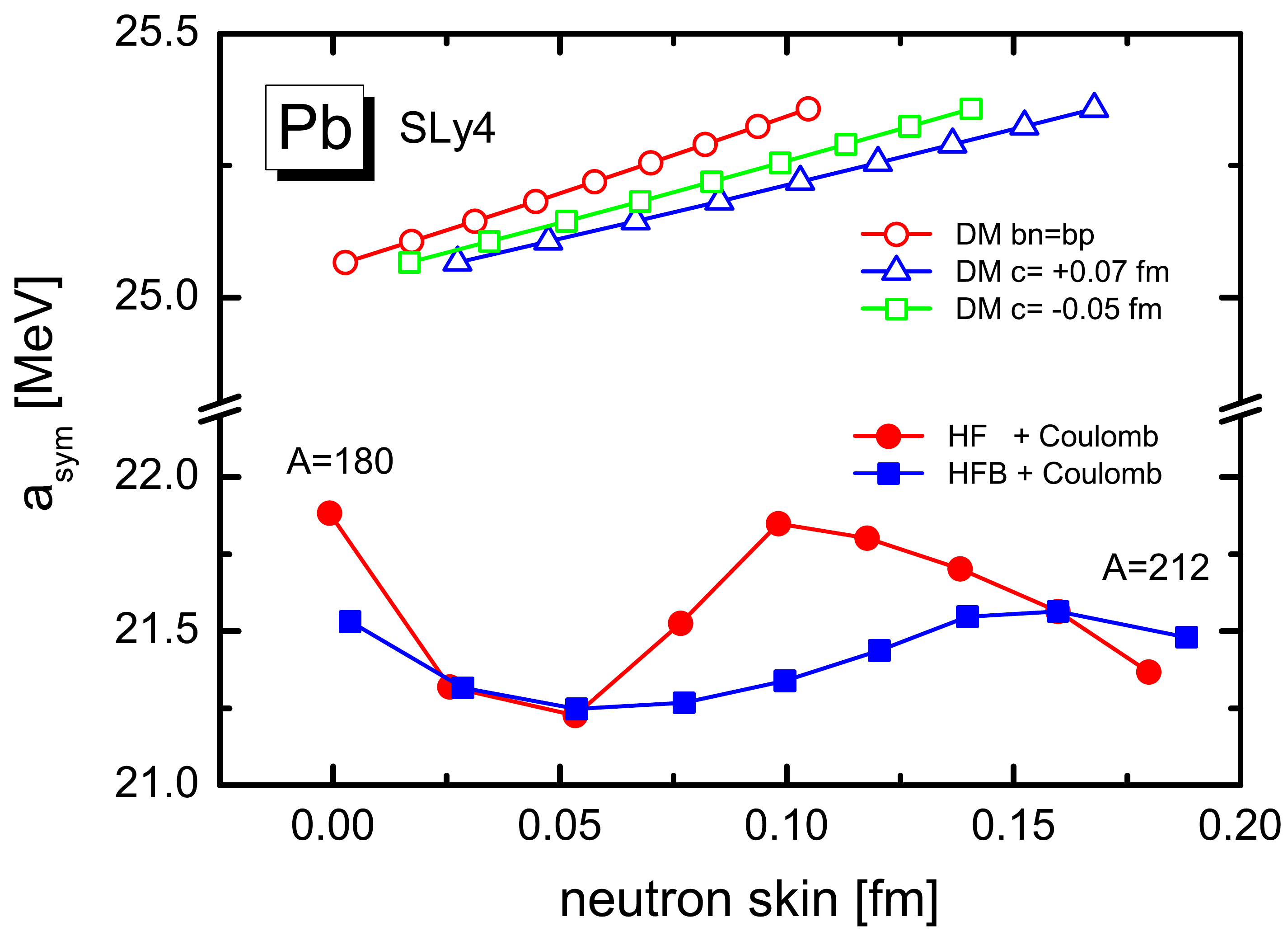}
	\caption{(Color online)  Symmetry energy $E_{sym}$ (upper panel) and coefficient $a_{sym}$ (lower panel) as a function of NST for Pb isotopes ($A = 180, 184, \cdots, 212$). The filled symbols denote the results given by the Skyrme functional SLy4 of Hartree-Fock-Bogoliubov (HFB) calculation with (square) and Hartree-Fock (HF) calculation without (circle)  pairing interaction. The open symbols denote the results in the droplet model (DM) without surface width corrections (circle), with the surface width correction of $c=-0.05$~fm (square) and $c=+0.07$~fm (triangle). }
	\label{fig2}
\end{figure}

\begin{figure}[h]
\includegraphics[width=0.85\linewidth]{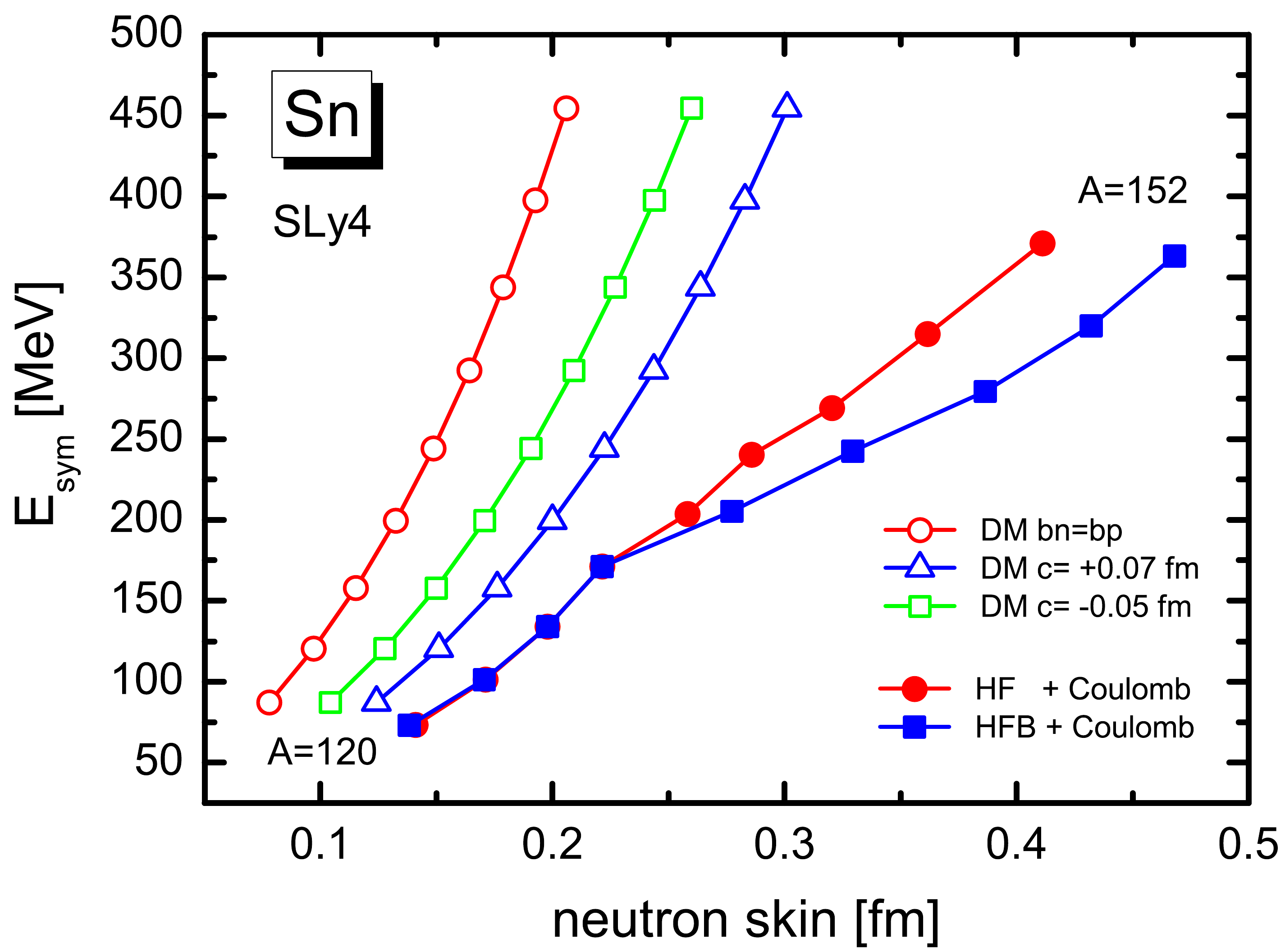}
\includegraphics[width=0.85\linewidth]{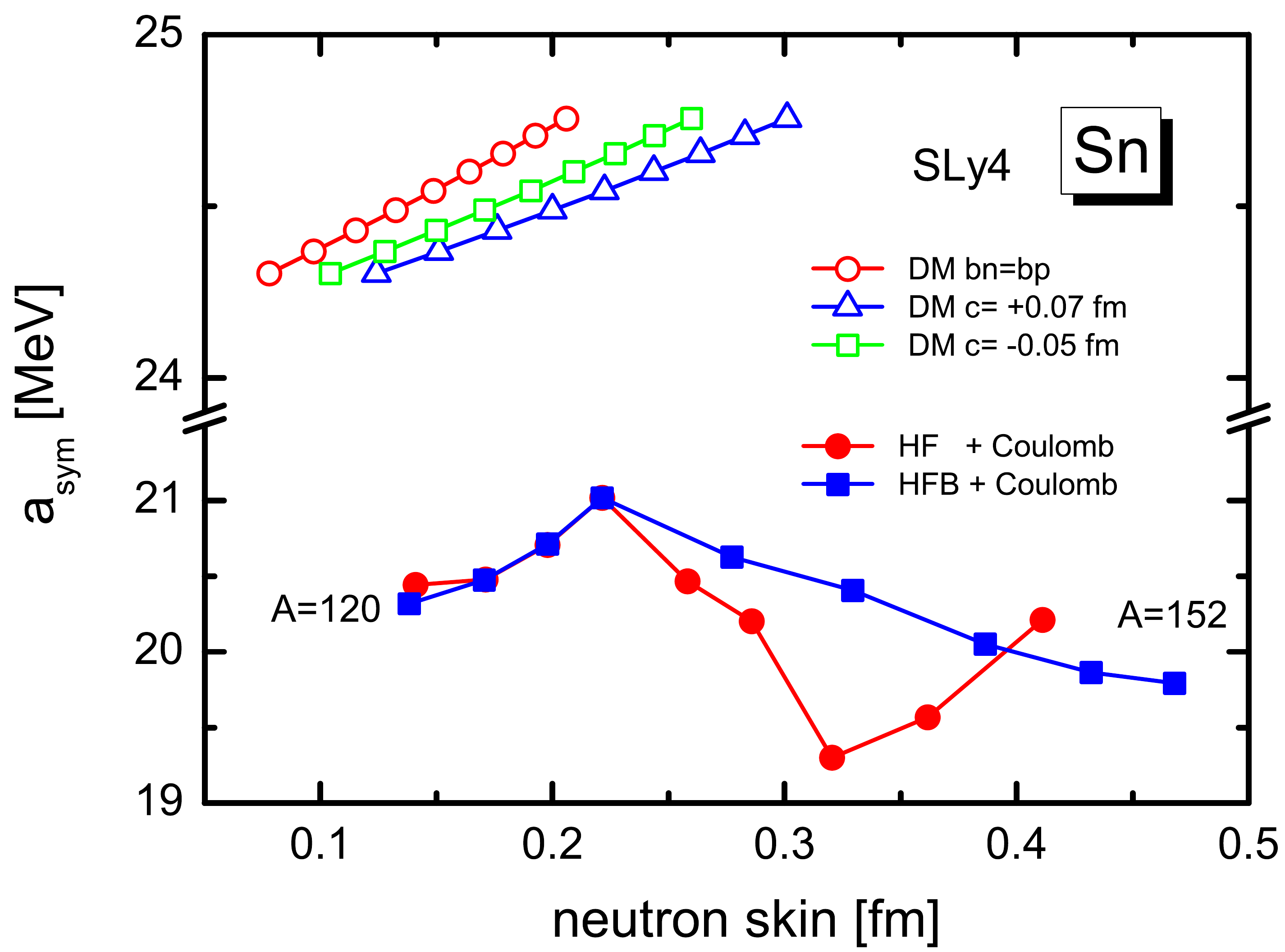}
\caption{(Color online) Similar as \fig{fig2}, but for Sn isotopes with $A=120, 124, \cdots, 152$.}
\label{fig3}
\end{figure}

Figures~\ref{fig2} and \ref{fig3} show the numerical results for the symmetry energy $E_{sym}(A)$ (upper panel) and the symmetry energy coefficient $a_{sym}(A)$ (lower panel)  as a function of the NST for isotopic chain Pb ($A=180, 184, \cdots, 212$) and Sn ($A=120, 124, \cdots, 152$), respectively, calculated with the Skyrme functional SLy4.
The filled circles denote the result of the Hartree-Fock results without pairing, while the filled squares stand for the result of the HFB results with the DDDI pairing.  For comparison, the results from the DM calculated by Eqs.\,\eqref{eq:EsymDM} and \eqref{eq:asymDM} are also shown in the figures by the open symbols, where the circles denote the NST without the surface width corrections, i.e., $b_n=b_p$ in Eq.~\eqref{eq:skin-sw-b}, and the triangles (squares) denote the results given by the parameterized surface width correction with $c=+0.07$~fm ($c=-0.05$~fm).  One should notice that, in our mean-field calculations, we can include self-consistently the shell and pairing effects, while the DM cannot.

Taking the Pb isotopes as examples, one can see from the upper panel in Fig.~\ref{fig2} a large increase of the symmetry energy $E_{sym}$ as a function of the NST for heavier nuclei for both the DM and mean-field calculations.
The NST is known to be proportional to the $L$ in the symmetry energy, which is fixed as $L$ = 100.53 for SLy4 interaction. If we may measure the NST, $S$, one can deduce the $E_{sym}$ for a given nucleus, or vice verse, from these results. Most of the symmetry energy given by the DM is larger than those given by the self-consistent mean-field calculations for a given isotope.  The surface width corrections for the NST becomes more and more important for the heavier nuclei in the DM model.

Only for the isotope $A=180$, the DM model without the surface width correction gives the similar results of the symmetry energy and the NST (almost zero) with the mean-field calculations.  For $A=184$, the NST of the mean-field calculation is closer to the DM result with $c=-0.05$~fm.  From $A=188$ to $A=204$, the NST given by the mean-field model lies between the surface width correction region $c=-0.05 \sim +0.07$~fm, which is consistent with the DM ansatz in Eq.~\eqref{eq:skin-sw-c}.

However, when we comes to the even more neutron-rich isotopes for $A=208, 212$, the mean-field calculation gives larger NST than DM.   At the same time, we notice that the pairing effect from the comparison between the HF and HFB calculations is almost negligible in this example.  This fact demonstrates that the larger NST in the mean-field calculations than the DM results is mainly due to the shell effects included in the HF mean field.

When we come to the symmetry energy coefficient $a_{sym}$ in the lower panel in Fig.~\ref{fig2}, the rate of change by the neutron skin is much smaller than that in $E_{sym}$ because the $A$ and $\delta$ dependence in the $E_{sym}$ disappears in the coefficient $a_{sym}$. In fact, the coefficient $a_{sym}$ turns out to be nearly independent of the NST, which means the coefficient $a_{sym}$ ( or $J$ ) is a property of nuclear matter and confirms the $L$ dependence of the NST. Also we can see more clearly that the results of DM is larger than those given by the mean-field models.

As expected from Eq.~\eqref{eq:tasym1}, we can see a linear increase of the symmetry energy coefficient as a function of the NST in the DM.  However, this relation is not found in the mean-field model at all. It means that the NST depends on the nuclear structure of each nucleus, ${\it i.e.}$ shell effects. Besides, the difference between the HF and HFB results in the lower panel is enlarged in the symmetry energy coefficient, $a_{sym}$ with a small scale although they give the similar NST. Namely, not only the shell effects but also the pairing correlations influence the $a_{sym}$ as well as the NST.  One can see the largest difference between in the $a_{sym}$ appears in the isotope $A=196$, which has $114$ neutrons and lies between the $82$ and $126$ major shells, while the smallest case appears at the double magic nucleus (A=208, N=126) as expected.

\begin{figure}[h]
\includegraphics[width=0.85\linewidth]{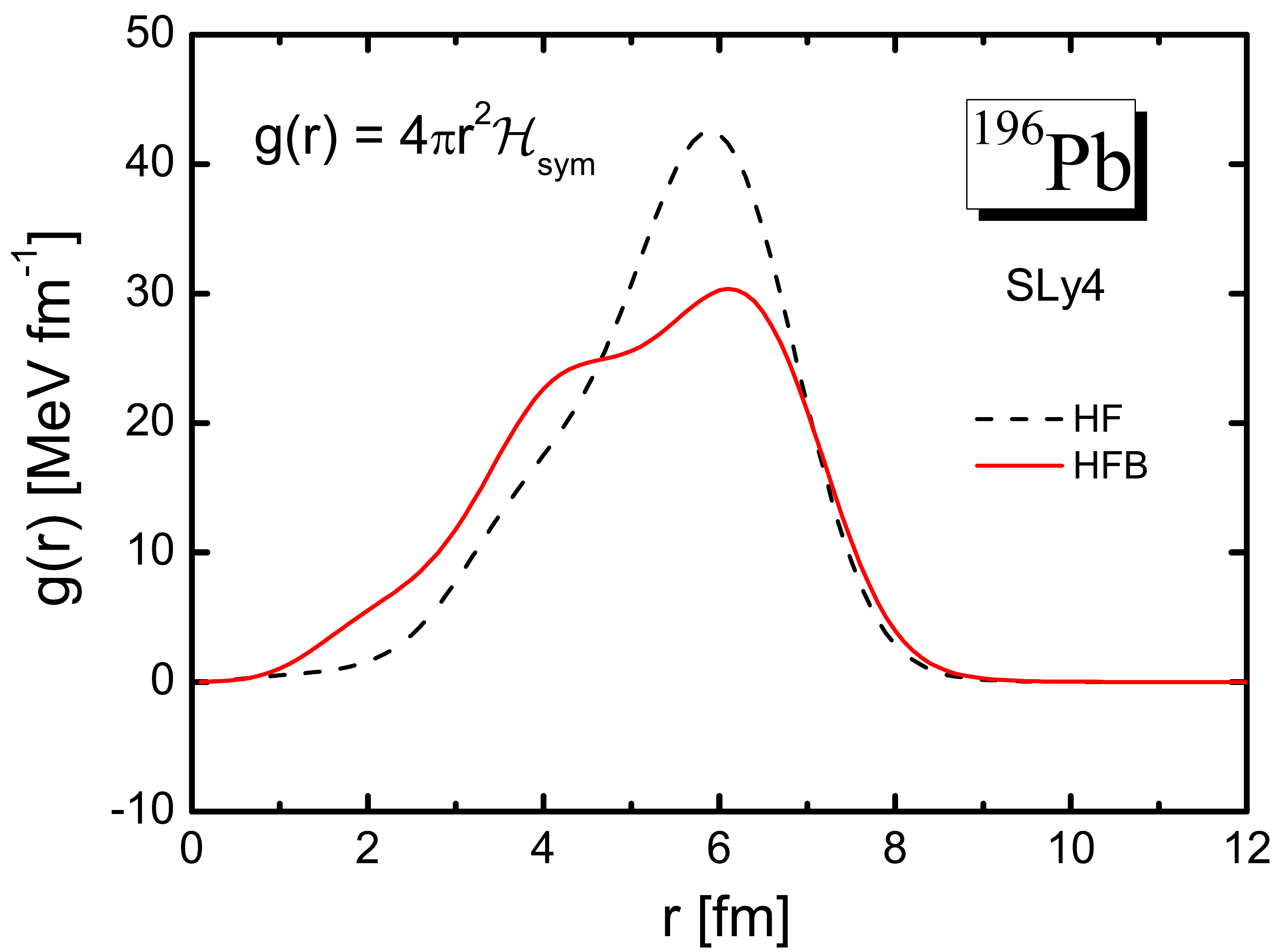}
\includegraphics[width=0.85\linewidth]{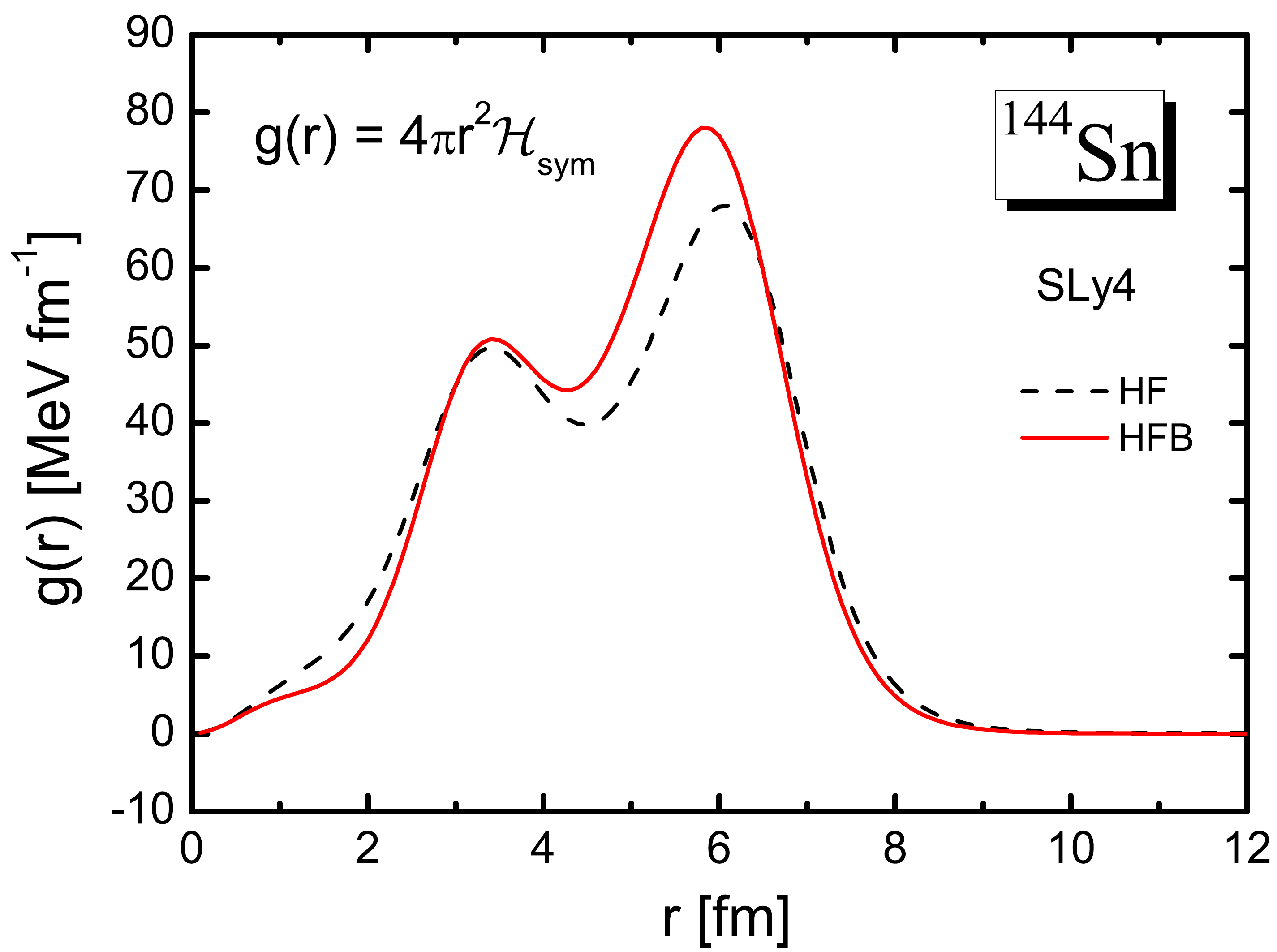}
\caption{(Color online) Symmetry energy density distribution $g(r)=4\pi r^2 \mathcal{H}_{sym}$ in \atom{Pb}{196} (upper panel) and \atom{Sn}{144} (lower panel) obtained by the HF(dashed line) and HFB (solid line) calculation with SLy4 functional. }
\label{fig4}
\end{figure}

In order to further understand the difference in the symmetry energy coefficient due to the pairing effect in $A=196$, we plot the symmetry energy density $g(r)=4\pi r^2\mathcal{H}_{sym}$ given by the HF (dashed line)  and HFB (solid line) calculations in the upper panel of Fig.~\ref{fig4}.  One can see that both the symmetry energy densities by HF and HFB are mainly contributed around the surface region $r\approx 6$~fm, where the HF result without pairing is more concentrated while the HFB result with pairing is more diffusive.  This gives a smaller symmetry energy coefficient with the pairing effect in this nucleus.

Figure~\ref{fig3} for Sn isotopes with $A=120, 124, \cdots, 152$ shows the similar calculation results with Fig.~\ref{fig2} for Pb isotopes.  First, one can see from the upper panel that the symmetry energy given by the DM is again systematically larger than that given by the mean-field model for a given isotope.  However, the NST of all these isotopes obtained by the mean-field model  is larger than those given by the DM, which is quite different from the Pb isotopes.  This demonstrates that the shell effects play an important role in producing a large NST in Sn isotopes.  From the comparison between the HF and HFB results, one can see that from $A=120$ to $A=132$, the pairing effects is almost negligible in NST.  However, from $A=136$ the pairing effects produce a larger NST for more neutron-rich nucleus, while a nucleus (A=132, N=82) has also no pairing interaction effects due to the double magic number.

From the lower panel in Fig.~\ref{fig3}, we can see again, in the HFB as well as HF mean field results, that there is no linear relation as the DM between the symmetry energy coefficient and the NST in these isotopes.  The largest difference between the HF and HFB results of the symmetry energy coefficient appears for the isotope $A=144$.  Therefore, we plot the symmetry energy density calculated with and without pairing effect in the lower panel of Fig.~\ref{fig4} in this isotope.  One can find that the pairing effect contributes more to the symmetry energy density and more concentrates around the surface region. This is consistent with the fact that the pairing effects become significant in the nuclei retaining wide smearing region as shown in other shell model and QRPA calculations \cite{Ha15,Ha16,Ha17}.

\begin{figure}[h]
\includegraphics[width=0.85\linewidth]{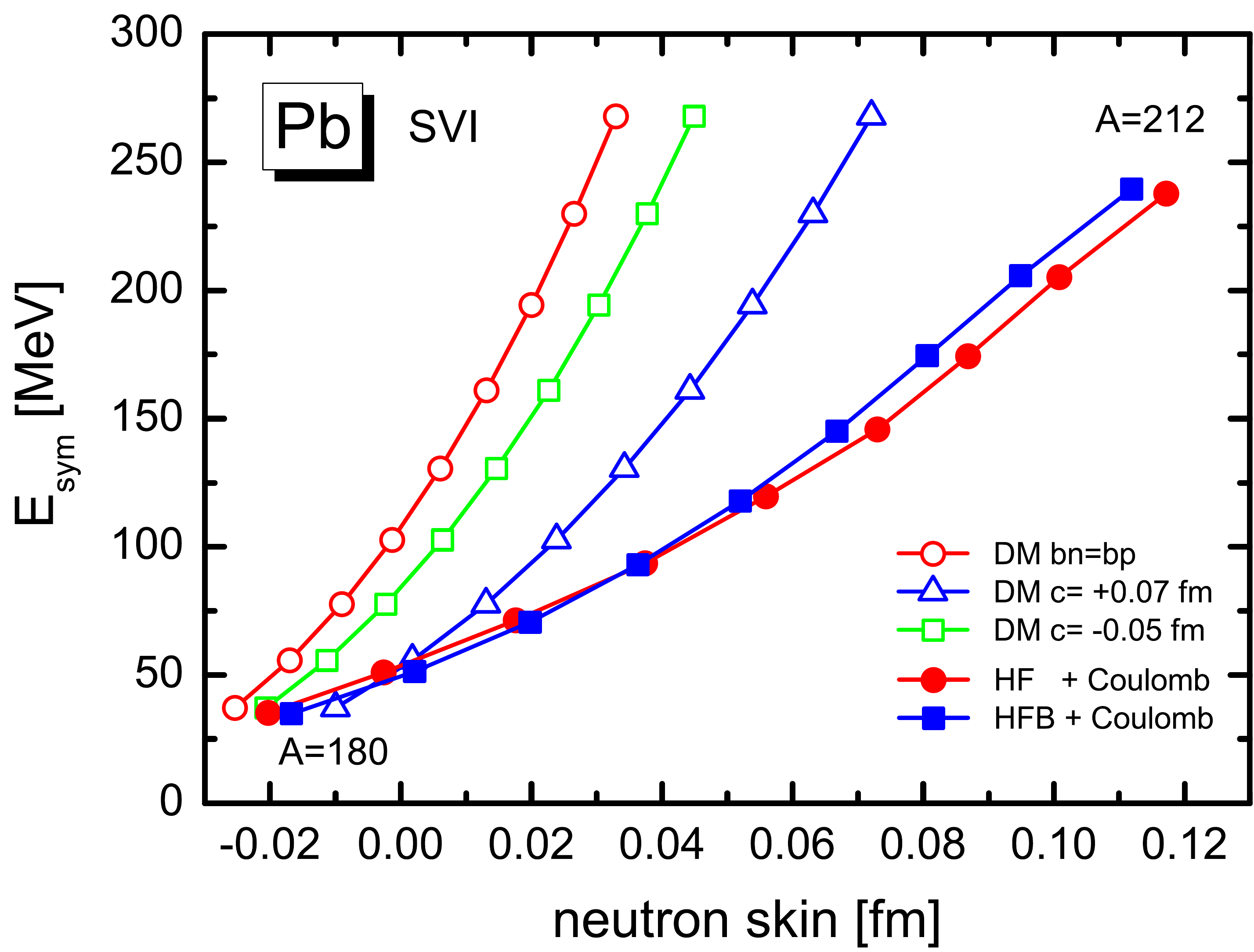}
\includegraphics[width=0.85\linewidth]{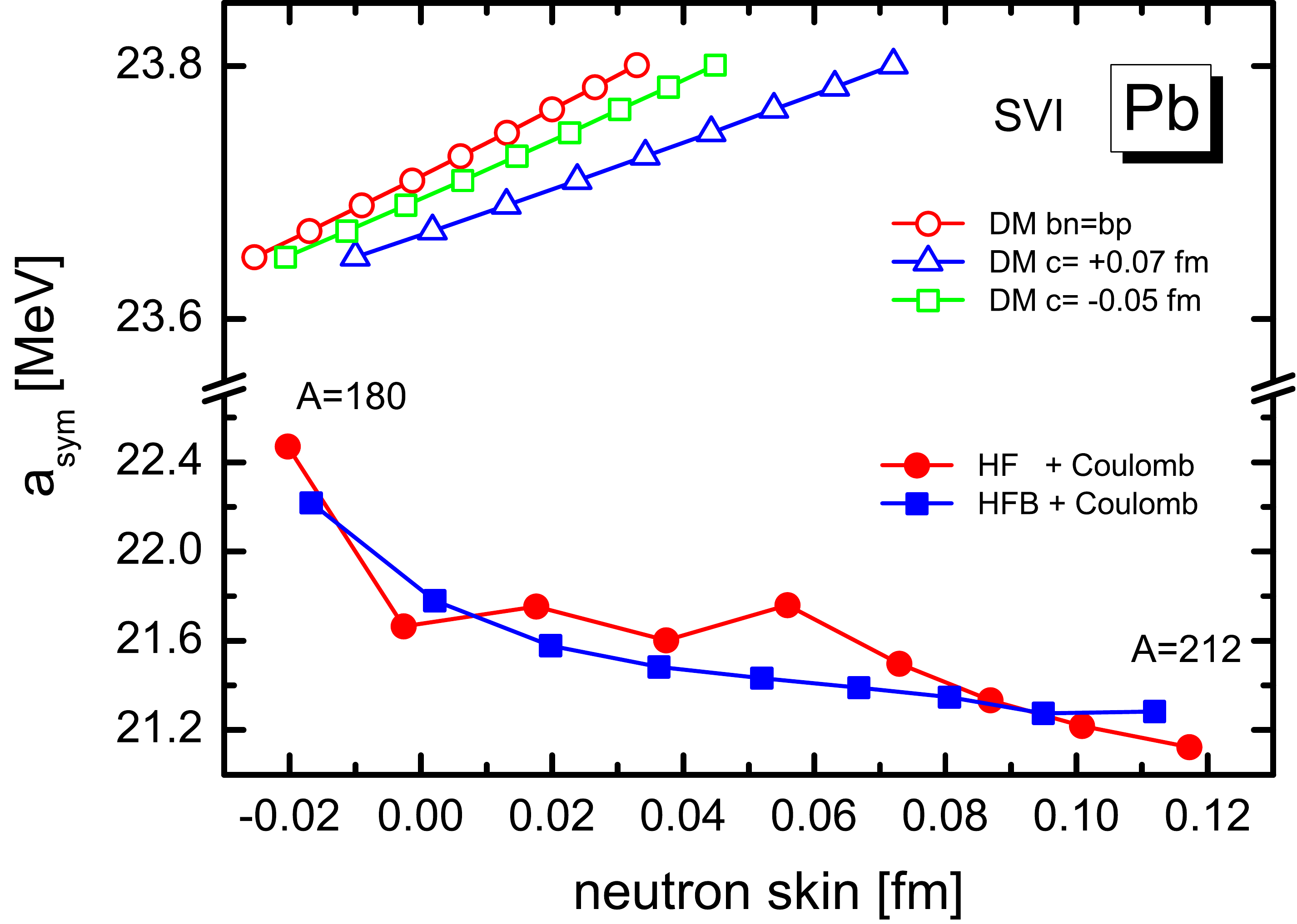}
\caption{(Color online)  Similar with~\fig{fig2}, but calculated by the Skyrme functional SVI.  }
\label{fig5}
\end{figure}

\begin{figure}[h]
\includegraphics[width=0.85\linewidth]{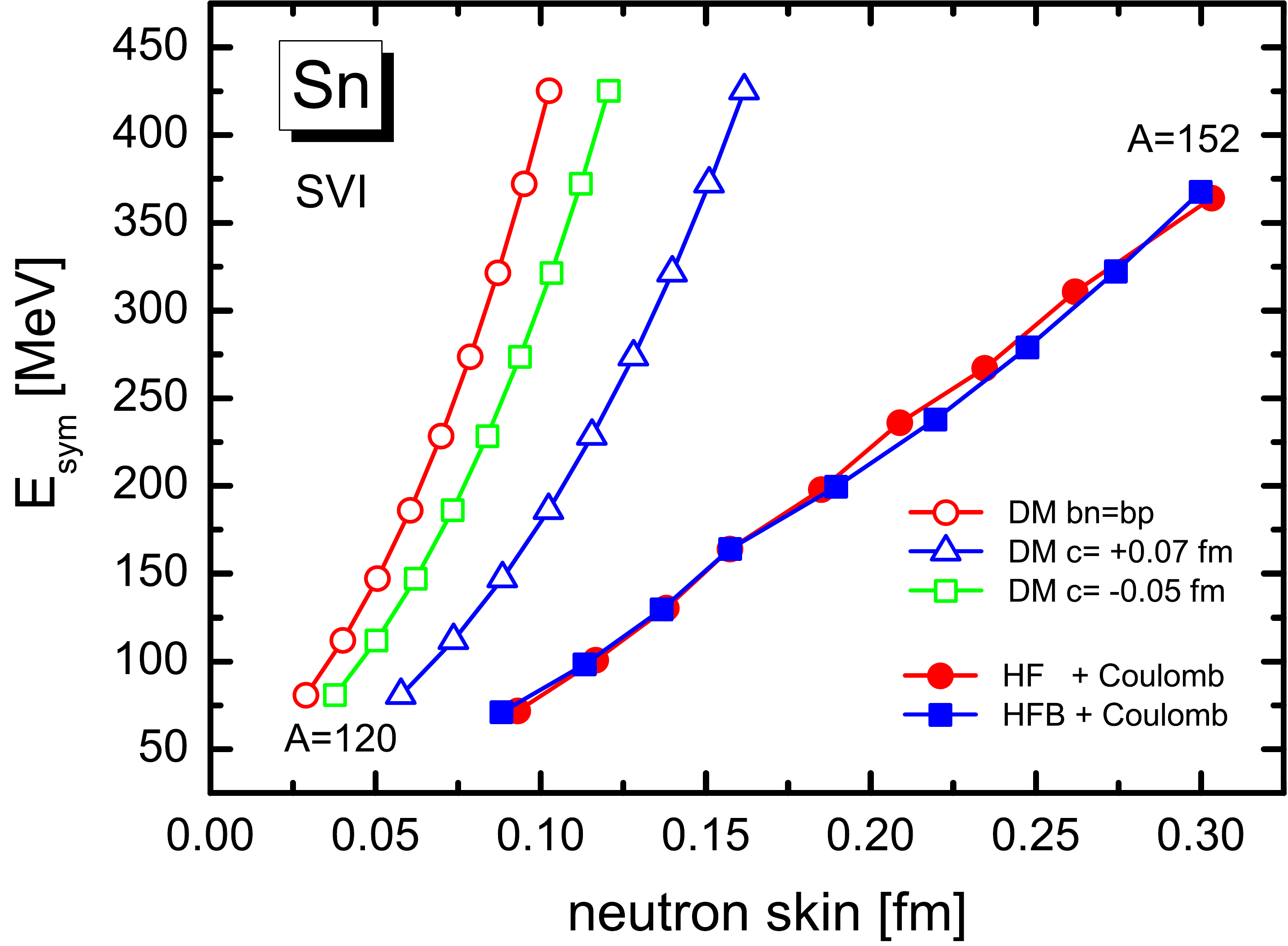}
\includegraphics[width=0.85\linewidth]{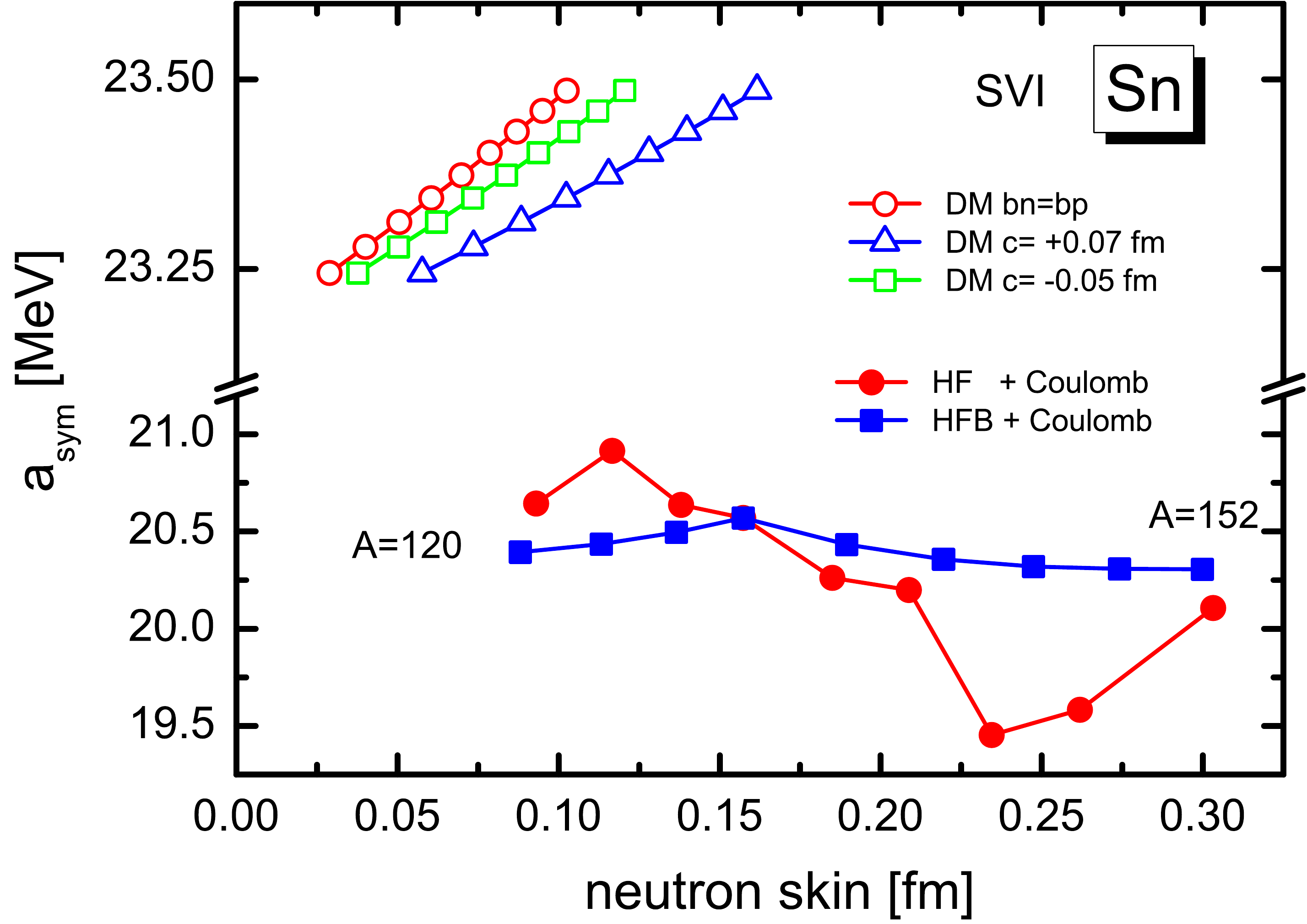}
\caption{(Color online) Similar with~\fig{fig3}, but calculated by the Skyrme functional SVI.}
\label{fig6}
\end{figure}

Similar analysis are done for Pb and Sn isotopes with SVI Skyrme functionals in Fig.~\ref{fig5} and~\ref{fig6}.  One can see that the overall behavior of the symmetry energy and coefficient are similar with those given by the SLy4.  But, the NST is much smaller than those by SLy4 because the $L$ value is much smaller than that of SLy4 functional. However, the pairing effects produce less different NST from the HF results in these two isotope chains with the SVI functional.

From the above, we can see that the mean-field models which include self-consistently the full shell and pairing effects gives less symmetry energy and coefficient, but larger NST, compared with the DM model which does not have these two effects.  In the mean-field model, the symmetry energy and coefficient are mainly contributed by the surface region, which is mainly driven by the shell and pairing effects.  In particular, the pairing effects play an important role in the symmetry energy and the NST in open-shell nuclei.


\section{Pairing in infinite matter}
\begin{figure}[b]
\includegraphics[width=0.85\linewidth]{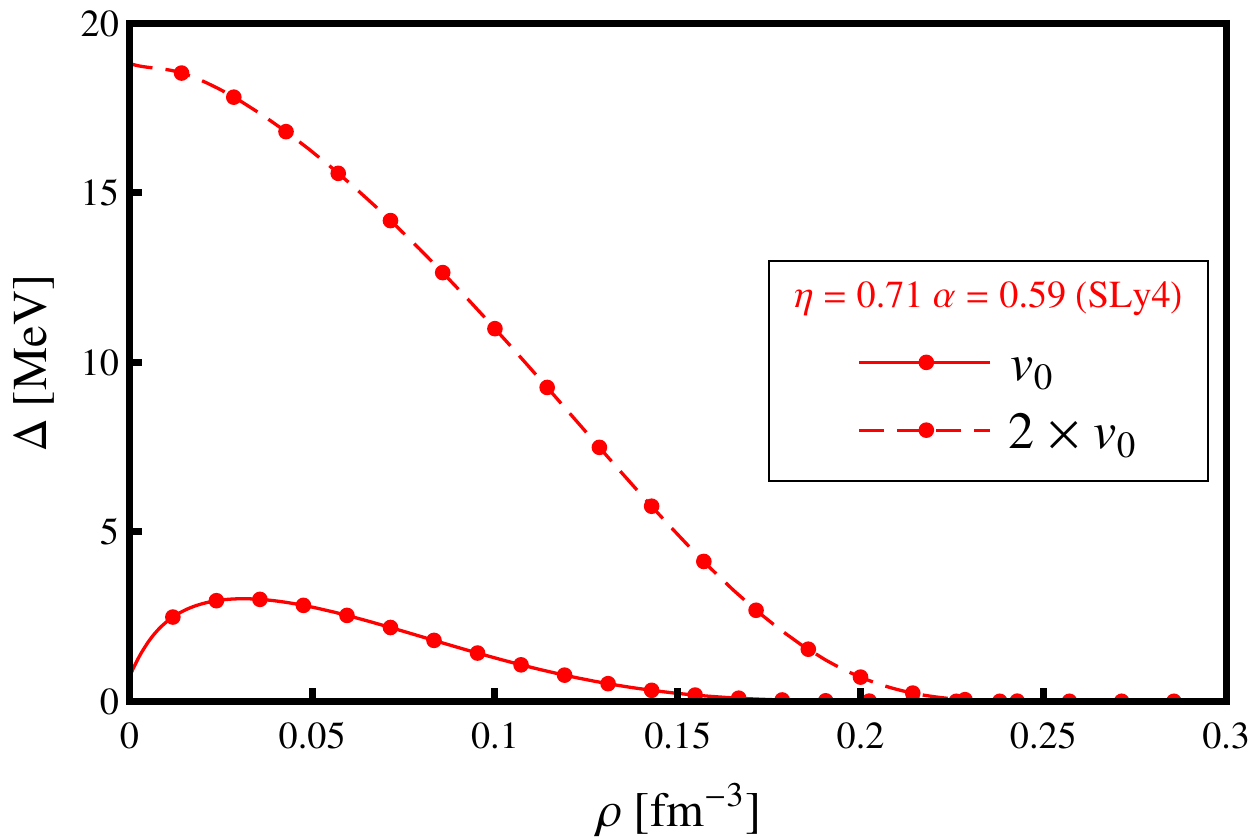}
\includegraphics[width=0.85\linewidth]{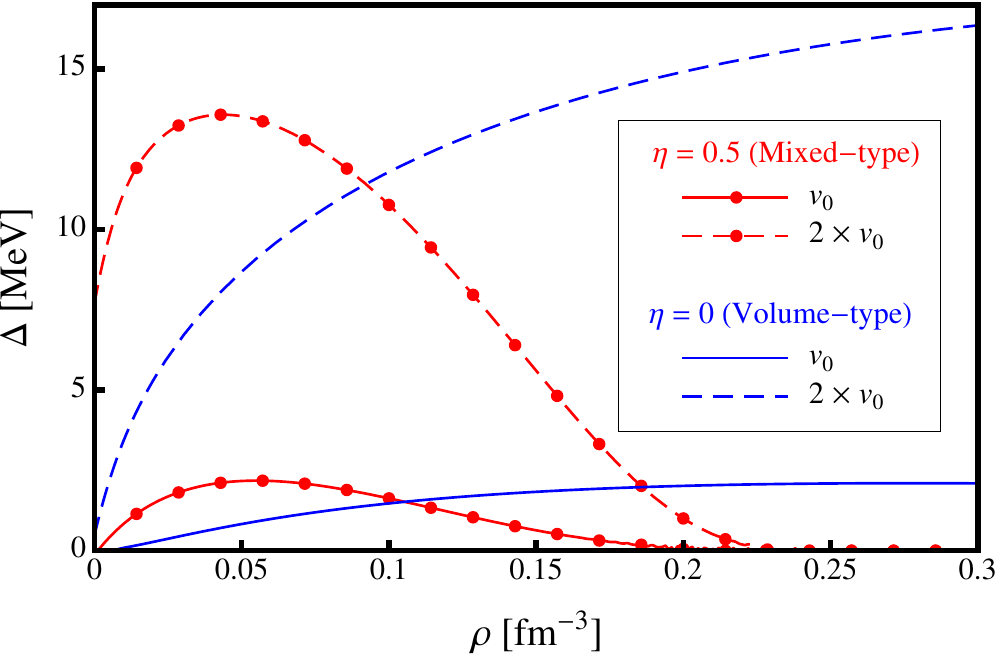}
\caption{(Color online) Pairing gap obtained with the density-dependent delta interaction of which parameters are determined in regard to the SLy4 (upper panel) and SLy5 Skyrme (lower panel) functionals.
}
\label{fig7}
\end{figure}

This section is devoted to the effects of the pairing correlations in infinite nuclear matter. Nuclear energy density for the nuclear matter can be written as a sum of Skyrme and pairing terms
\begin{equation}
	\mathcal{E} = \mathcal{E}^{}_\text{Skyrme} + \mathcal{E}^{}_\text{pair}~.
	\label{eq16}
\end{equation}
The Skyrme energy density, $\mathcal{E}^{}_\text{Skyrme}$, corresponds to \eqn{eq10}, but the Coulomb, spin-orbit, and gradient terms are omitted for the case of infinite matter.
In the BCS approximation as the simplest approach, the pairing energy density is defined by
\begin{equation}
	\mathcal{E}^{}_\text{pair} = -\frac{1}{2}\sum_{\tau=n,\,p}N^{}_\tau \Delta^2_\tau~,
	\label{eq17}
\end{equation}
where $N^{}_\tau=k^{}_{F,\tau} m^\ast_\tau  / \pi^2 \hbar^2$ is the density of neutron ($\tau=n$) and proton ($\tau=p$) states, and the pairing gap, $\Delta^{}_\tau$, can be obtained by solving the BCS gap equation,
\begin{equation}
	\Delta^{}_\tau = \frac{G^{}_\tau}{2}\int^{k^{}_{F,\tau}}_0 \frac{\Delta^{}_\tau}{\sqrt{(\epsilon(k)^{}_\tau-\lambda)^2+\Delta^2_\tau}} \,\ud k^3~.
	\label{eq18}
\end{equation}
The denominator of the integrand in \eqn{eq18} denotes the quasi-particle energy with the chemical potential, $\lambda$, and the single-particle energy, $\epsilon(k) = \hbar^2 k^2/2m^\ast$, which is assumed to be the energy of free particle with the effective mass, $m^*$, in nuclear matter. First, we use SLy4 functional with the density-dependent contact interaction, \eqn{eq15}, for the pairing forces used in the previous section for finite nuclei. Second, we take SLy5 functional with the pairing field, $G^{}_\tau$, written in the form of
\begin{equation}
	G^{}_\tau = v^{}_0 \left[1-\eta\left(\frac{\rho^{}_\tau}{\rho^{}_0}\right)^{\alpha\,}\right]~.
	\label{eq19}
{\tiny }\end{equation}
For nuclear matter by the SLy5, we have used two types of pairing interaction adjusting the parameter, $\eta=0.5$ for mixed-type and $\eta=0$ for volume-type interaction. The strength parameter $v^{}_0$ for each $\eta$ is determined to reproduce the pairing gap of \atom{Sn}{120} ($\Delta\simeq1.321\MeV$) obtained by the HFB calculation with SLy5~\cite{Cao:2012dt}. For SLy5, the energy cutoff for the pairing window is taken to be $60\MeV$ and
\begin{align}
	v^{}_0 = \left\{ \begin{array}{ll}
 		-218\MeV fm^{3} & \text{for $\eta=0$} \\
		-325\MeV fm^{3}& \text{for $\eta=0.5$}~.\\
  \end{array} \right.
  \label{eq20}
\end{align}
For simplicity, we fix the parameter $\alpha$ to be one since the pairing strength is rarely sensitive to it~\cite{Dobaczewski:2001ed}.

The pairing gaps obtained by self-consistently solving \eqn{eq18} are shown for SLy4 (upper panel) and SLy5 (lower panel) functionals in \fig{fig7}. The density-dependence in the mixed-type interaction in SLy5 lowers the pairing gap at high densities because of the surface-type pairing, similarly to those by SLy4 which is also a kind of mixed type interaction due to non-zero $\eta$ value. In contrast to the low density region where the pairing gaps are microscopically understood, there are still ambiguities at higher densities than the nuclear saturation density. Such a ultra-dense nuclear matter is thought to acquire superfluidity through the formation of nucleon Cooper pairs due to the dominance of the long-range attractive part of the nucleon-nucleon ($NN$) interaction. In this context, it may be interesting what if the strength $v^{}_0$ is far from the values constrained by finite nuclei. Then we have also tested by multiplying a factor 2 to the strength in \eqn{eq15} and \eqn{eq20} for SLy4 and SLy5, respectively.

Moreover, recently, there are many discussions regarding the pairing correlations by the $T$=0 channel contribution coming from the unlike-pairing by the neutron-proton ($np$) pairing correlations \cite{Ha16,Bertch11,Sagawa16,Frau14}. The $np$ paring has two components $T$=0 and 1, while the like-pairing correlations from neutron-neutron and proton-proton have only $T$=1 contribution. In particular, the $T$=0 channel was shown to have pairing gaps much larger, about twice, than $T$=1 pairing channel \cite{Gar99,Gar01}. Here we tested the $T$=0 pairing contribution on the nuclear matter case by adopting the factor 2, whose results are presented as dashed lines in \fig{fig7}.

\begin{figure}[t]
\includegraphics[width=0.70\linewidth]{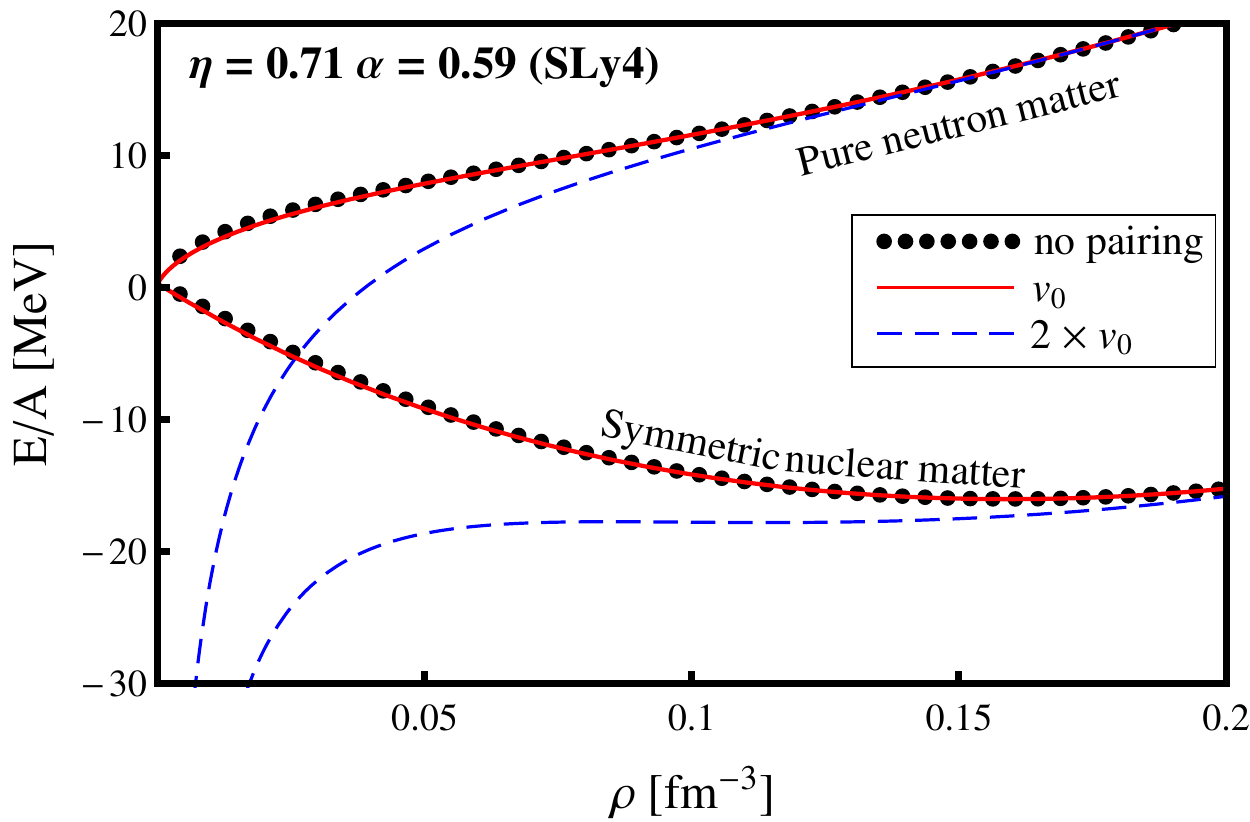}
\includegraphics[width=0.70\linewidth]{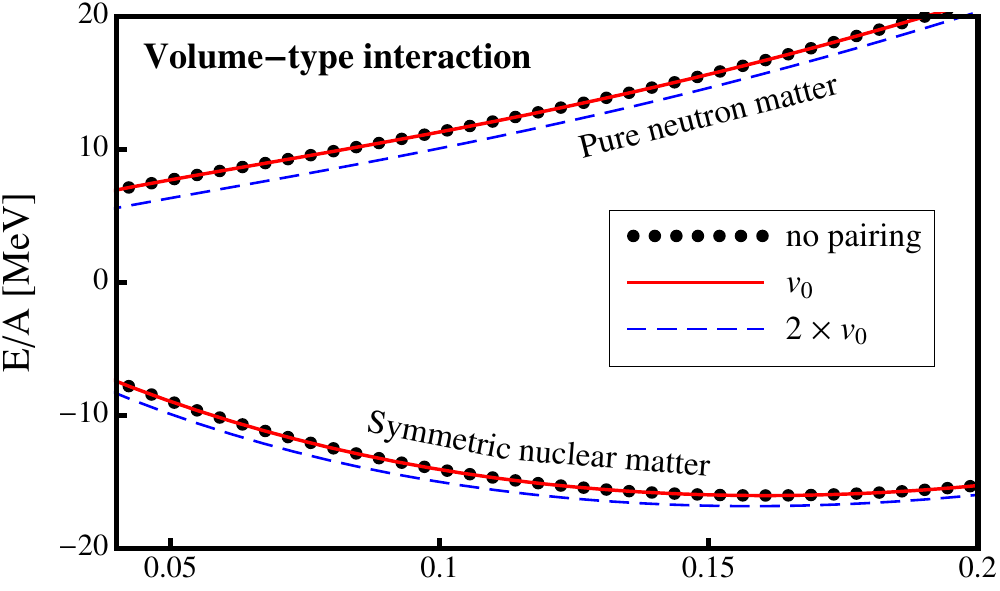}
\includegraphics[width=0.70\linewidth]{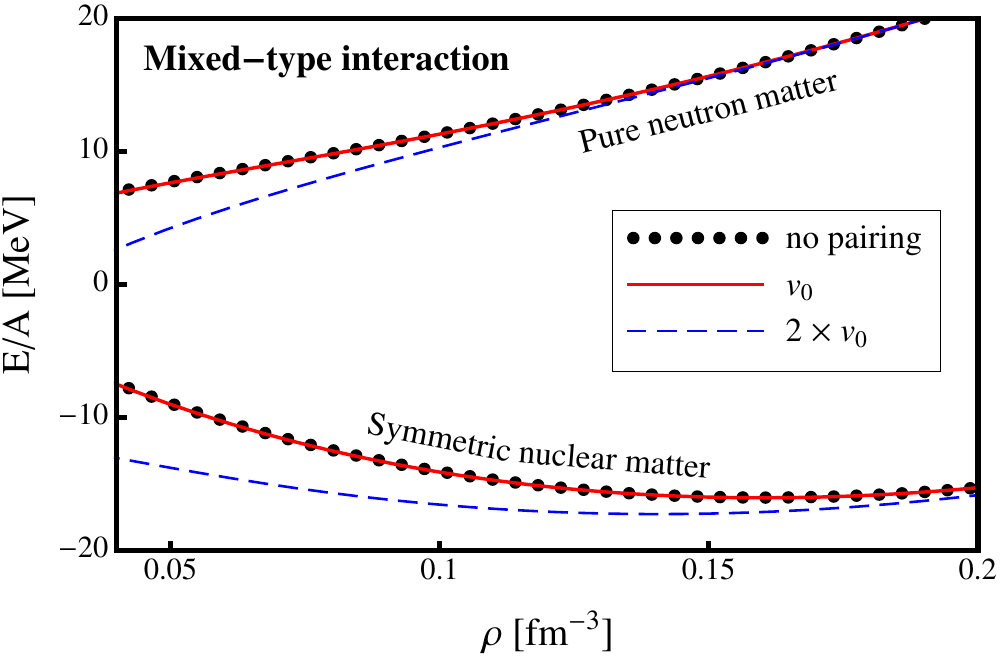}
\caption{(Color online) Energy per nucleon versus density in symmetric nuclear and pure neutron matter for the SLy4 (upper panel) and SLy5 Skyrme (lower two panels) functionals.}
\label{fig8}
\end{figure}

Finally, by using the energy functional in \eqn{eq16}, we calculate the EoS of the system, i.e. the energy per nucleon as a function of density. The EoS of the pure neutron  matter and also the symmetric nuclear matter in the BCS approximation are shown in \fig{fig8}. One may notice that, in the difference between dotted and solid lines, the pairing effects by the strength $v_0$ make only a difference less than a few electron-volt (eV) in the energy per nucleon, irrespective of the kinds of the Skyrme functionals. Next we increase the strength parameter $\eta$ twice by the $T$=0 channel, then the $E/A$ becomes lower in the lower density region. Specifically, the change in SLy4 functional makes the effect significant. But, in the high density region, the enlarged pairing gap effects appear only in the volume type interaction of the SLy5. Systematic calculations involving the microscopic effects, e.g. three-body forces, detailed $T$=0 channel pairing effect, {\it etc.}, are required to properly describe the pairing in the nuclear matter.

%

%
%

\section{Summary and Conclusions}


We calculated the neutron skin thickness (NST) of heavy nuclei by using a droplet model and three Skyrme functionals SLy4, SkI3 and SVI, which have quite different energy dependence, characterized by the slope $L$ and the curvature $K_{sym}$ in the symmetry energy. A linear relation between the NST and the asymmetry parameter $\delta$ can be deduced from the antiprotonic atom x-ray measurement data. Another linear relation between the NST and the $L$ has been widely discussed. We compare our results of NST with those experimental data, which are consistent within the error bars. But it turns out to be still difficult to extract the information of the slope $L$ from this investigation with the present experiment accuracy.
Namely, the three different Skyrme functionals retaining different $L$ values could reproduce the data within experimental error bars.

We also calculated the symmetry energy and its coefficient in terms of the NST  for Pb and Sn isotopes from the self-consistent Skyrme models and compared to those given by the droplet model.  We found that most of the mean-field calculations give less symmetry energy but larger NST than those by the droplet model. The pairing effects as well as the shell structure are shown to affect more or less the neutron skin thickness. In order to argue the effects from the pairing correlation, we also compare the results between the Hartree-Fock and Hartree-Fock-Bogoliubov calculations. It turned out that the symmetry energy density is mainly distributed around the nuclear surface region and the pairing effects also shows up mainly on the surface region in the symmetry energy and the NST for some open-shell nuclei having wide smearing.


By using the same Skyrme functional SLy4 used for finite nuclei, we calculated the pairing effects on the nuclear matter. Results by SLy4 and mixed type SLy5 functionals show maximally about 2.5 MeV pairing gaps in the low density region. But the volume type of the pairing by SLy5 leads to about 2.5 MeV pairing gaps, which are saturated even above the normal density region. If we take into account the enlarged $v_0$ strength assumed to come from the $T=0$ pairing channel, the pairing gaps were increased up to 13 $\sim$ 19 MeV by SLy5 and SLy4 types in the low density region, respectively. But the volume type SLy5 shows about 17 MeV pairing gap in the high density region.
In the binding energy, as shown in \fig{fig8}, the pairing effects were not so salient for SLy5, but becomes stronger by the enlarged $T=0$ pairing correlation channel. More stronger binding energy by this enlarged $T=0$ pairing contribution was also confirmed for SLy4 interactions.


In conclusion, first, our results by different Skyrme functionals show that the slope of the symmetry energy, $L$, is a key factor for determining the NST. For example, the NST by different Skyrme functionals, which have different $L$ values, reveals large differences as shown in the x-axis in Figs. 2-3, and 5-6. Second, for a given symmetry energy, the NST by DM is smaller than Skyrme functionals, which comes from the shell effects considered in the mean-field calculations.
Third, the pairing effects as well as the shell effects may play meaningful roles in the NST, but they are shown to be subsidiary to a role of the $L$ in the symmetry energy. Nevertheless they turned out to contribute to the symmetry energy and the asymmetry coefficients. Finally, the $T=0$ pairing contribution should be reexamined for proper understanding the nuclear matter.
Future experiments for the neutron skin thickness of Pb and Ca may deduce the precise constraint for the symmetry energy from those experiments. In particular, we need more detailed experimental data and more refined calculations of $T=0$ contribution by Skyrme functional approach for further conclusion of the pairing effects in the nuclear matter.

\section*{ACKNOWLEDGEMENTS}
This work was supported by Korea National Research Foundation under grant numbers Grant Nos. NRF-2014R1A1A1038328, NRF-2014R1A2A2A05003548 and NRF-2015K2A9A1A06046598, and the National Natural Science Foundation of China under No.
11405116.


\end{document}